# Are Weakly Coordinating Anions Really the Holy Grail of Ternary Solid Polymer Electrolytes Plasticized by Ionic Liquids? Coordinating Anions to the Rescue of the Lithium Ion Mobility

[1]Jan-Philipp Hoffknecht[a,b], [1]Alina Wettstein[c], Jaschar Atik[d], Christian Krause[b], Johannes Thienenkamp[d], Gunther Brunklaus[d], Martin Winter[b,d], Diddo Diddens[d(*)], Andreas Heuer[c,d(*)], Elie Paillard[d,e*]

[a]*Institute for Inorganic and Analytical Chemistry, University of Münster, Corrensstrasse 28/30, D-48149 Münster, Germany*

[b]*MEET Battery Research Center, University of Münster, Corrensstrasse 46, D-48149 Münster, Germany*

[c]*Institute for Physical Chemistry, University of Münster, Corrensstrasse 28/30, D-48149 Münster, Germany*

[d]*Helmholtz Institute Münster – Forschungszentrum Jülich GmbH (IEK 12), Corrensstrasse 46, D-48149 Münster, Germany*

[e]*Politecnico di Milano, Dept. Energy, via Lambruschini 4, 20148 Milan, Italy*

[1]These authors contributed equally to this work.

Abstract:

Lithium salts with low coordinating anions like bis(trifluoromethanesulfonyl)imide (TFSI) have been the state-of-the-art for PEO-based 'dry' polymer electrolytes for three decades. Plasticizing PEO with TFSI-based ionic liquids (ILs) to form ternary solid polymer electrolytes (TSPEs) increases conductivity and Li$^+$ diffusivity. However, the Li$^+$ transport mechanism is unaffected compared to their 'dry' counterpart and essentially coupled to the dynamics of the polymer host matrix, which limits Li$^+$ transport improvement. Thus, a paradigm shift is hereby suggested: The utilization of more coordinating anions such as trifluoromethanesulfonyl-*N*-cyanoamide (TFSAM), able to compete with PEO for Li$^+$ solvation to accelerate the Li$^+$ transport and reach higher Li$^+$ transference number. The Li–TFSAM interaction in binary and ternary TFSAM-based electrolytes was probed by experimental methods and discussed in the context of recent computational results. In PEO-based TSPEs, TFSAM drastically accelerates the Li$^+$ transport (increased Li$^+$ transference number by 600% and Li$^+$ conductivity by 200-300%) and computer simulations reveal that lithium dynamics are effectively re-coupled from polymer to anion dynamics. Finally, this concept of coordinating anions in TSPEs was successfully applied in LFP‖Li metal cells leading to enhanced capacity retention (86% after 300 cycles) and an improved rate performance at 2C.

*Corresponding author: E.P.: elieelisee.paillard@polimi.it
(*)Corresponding authors (computational methods): wettstein@uni-muenster.de

## 1 Introduction

Polyethylene oxide (PEO)-based electrolytes are the current state-of-the-art 'dry' solid polymer electrolytes (SPEs).[1] They reach ionic conductivities around $10^{-4}$ S cm$^{-1}$ at 40°C, thanks to the strategies that have been developed to suppress the crystallinity of linear PEOs and their crystalline complexes (*i.e.*, crosslinking, statistical copolymers and 'plasticizing' lithium salts).[2] However, for an application in automotive batteries, conductivities around $10^{-3}$ S cm$^{-1}$ and higher are necessary which implies an operational temperature above 60 °C for the present level of development of lithium metal polymer batteries (LMPBs). One strategy for decreasing the operation temperature of SPEs has been the use of plasticizers yielding ternary SPEs (TSPEs), for which both the segmental mobility of the PEO host and the transport properties of the ionic species are increased.[3,4] Among all the proposed compounds, ionic liquids (ILs) present the significant advantage of being non-volatile and non-flammable in most circumstances.[5–8] A common design rule is to combine a low coordinating anion with an organic cation, such as a tetra-alkyl ammoniums. Low coordinating anions such as $PF_6^-$, bis(trifluoromethanesulfonyl)imide (TFSI) or bis(fluorosulfonyl)imide (FSI) are typically used in lithium-ion and lithium metal battery electrolytes. In fact, organic and very weakly coordinating anions such as TFSI were first proposed for 'dry' SPEs.[9] Indeed, although PEO possesses a high donor number which favors the solvation of lithium and thereby promotes the solubility of lithium salts, its dielectric constant is relatively low. Hence, low coordinating anions such as TFSI result in easily soluble, low lattice energy salts that reach the highest dissociation and the best conductivities in 'dry' SPEs. However, these anions neither interact much with the PEO matrix nor Li$^+$, in regard of which they do not compete with the strong Li$^+$ coordination ability of PEO. As a result, their mobility is much higher than that of the Li$^+$ cation, which results in lithium transference numbers ($t_{Li^+}$) much lower than those typically observed in liquid organic electrolytes, which is detrimental for the battery performance.[10,11] For instance, PEO/LiTFSI complexes achieve $t_{Li^+}$ of about 0.1-0.15[12,13] as opposed to around 0.3-0.4 for liquid, alkyl carbonate-based electrolytes.[14] Given that TFSI is a large and weakly coordinating anion with a high conformational flexibility[15,16] that does not promote the crystallization of ILs or PEO-based electrolytes and is highly resistant to hydrolysis, it allows preparing stable ILs with low melting points and theses ILs, in turn, allow preparing ternary solid polymer electrolytes with considerably improved conductivity at 40°C (up to $10^{-3}$ S cm$^{-1}$).[17] Nevertheless, the introduction of extra ions in the form of ILs results in a further decrease of $t_{Li^+}$, which is obstructive for reaching high steady-state currents. In fact, it has been shown that tetra alkyl ammonium TFSI-based ILs such as *N*-butyl-*N*-methyl pyrrolidinium (PYR$_{14}$) TFSI act strictly as a plasticizer and that their effect is linked to an acceleration of PEO-related conduction modes, since they do not compete significantly with PEO for Li$^+$ solvation.[18] Considering that, for different ILs with the same cation, viscosity and conductivity are more linked to the size of the anion than to its coordinating ability,[19] it seems promising to use ILs with lower viscosity to reach a higher plasticizing effect.

Besides, it can be advantageous to introduce other coordinating groups into the system in order to enable additional conduction pathways for Li$^+$. For instance, we showed recently that, by use of ILs with a coordinating cation (*i.e.*, a TFSI-based IL with a pyrrolidinium cation bearing a short PEO chain long enough to solvate a single Li$^+$ cation), it was possible to triple the transference number of TSPEs compared to an alkyl-substituted IL counterpart.[13,20] However, it must be pointed out that the IL is extremely viscous. Moreover a part of its plasticizing effect at a given molar fraction originates from its very high molecular weight (given that the $T_g$ of a mixture of two compounds a and b scales with mass fractions according to the Flory-Fox law (*i.e.*, $1/T_{g,ab}$= $x_a/T_{g,a} + x_b/T_{g,b}$), with $x_a$ and $x_b$, the mass fractions of each compounds).[21]

Departing from the strategy of coordinating cations, we report here an approach based on solvating anions, employing a recently reported IL made of an asymmetric anion that can effectively prevent crystallization, recently proposed for IL-based electrolytes, trifluoromethanesulfonyl-*N*-cyanoamide (TFSAM). This anion is a hybrid between TFSI and dicyanoamide (DCA), as its negative charge is delocalized on one side within the -SO$_2$CF$_3$ strongly electron-withdrawing group, and on the other side, by a -CN group. Compared to the SO$_2$CF$_3$, the -CN group is not only slightly less electronegative but more polar and more coordinating itself. This anion, which is stable to hydrolysis, is asymmetric, which allows obtaining fully amorphous ILs and Li$^+$ electrolytes. Its anodic stability, although slightly lower than that of TFSI, is sufficient for the application and, contrary to TFSI, it presents the added benefit of not inducing any anodic dissolution of aluminum current collector at high voltage, even in carbonate-based lithium-ion battery electrolytes.[22,23]

We show here in a combined experimental and computational study that, contrary to the common approach of using ILs that comprise the most weakly coordinating anions which perform best in 'dry' binary SPEs, using a smaller and more coordinating anion is beneficial in the framework of TSPEs. Recently, we suspected a similar effect (although to a lesser extent) induced by the fluorine-free 4,5-dicyano-1,2,3-triazolat (DCTA) anion. However, in this case, it could only partially counterbalance the much lower conductivity *vs*. the TFSI analog and Li metal cells had a lower performance.[24] With TFSAM, the new 'anion-assisted' Li$^+$ transport mechanism allows a drastically faster Li$^+$ transport in TSPEs compared to the TFSI analog. This leads to an improved rate performance of LMPBs and an enhanced capacity retention of LMPBs in the long term as well.

## 2 Experimental

### 2.1 Materials

The following materials were used in this study: Polyethylene oxide (PEO, 5M, Sigma Aldrich), lithium bis(trifluoromethanesulfonyl)imide (LiTFSI, Sigma-Aldrich, 99.95%), lithium (2,2,2-trifluoromethane-sulfonyl)-*N*-cyanoamid (LiTFSAM, Provisco CS, 99%), *N*-butyl-*N*-methyl pyrrolidinium bis(trifluoromethanesulfonyl)imide (Pyr$_{14}$TFSI, Solvionic SA, 99.9%), *N*-butyl-*N*-methyl pyrrolidinium dicyanamide (Pyr$_{14}$DCA, Solvionic SA, 99.9%), lithium iron phosphate (LFP, Südchemie), polyvinylidene fluoride (PvdF 5140, Solef), Super P carbon black (Imerys), lithium metal foil (Albemarle, batterie grade, thickness: 50 µm).

*N*-butyl-*N*-methyl-pyrrolidium (2,2,2-trifluoromethyl-sulfonyl)-*N*-cyanoamide (Pyr$_{14}$TFSAM) and lithium dicyanamide (LiDCA) were prepared and purified as published earlier.[23]

### 2.2 Electrolyte preparation

All components were dried under high vacuum (10$^{-7}$ mbar) and respectively (LiTFSI: 100 °C for 3 d; LiTFSAM and LiDCA: 80 °C, for 5 d; Pyr$_{14}$TFSI: 100 °C, 5 d; Pyr$_{14}$TFSAM and Pyr$_{14}$DCA: 80 °C, 7 d; PEO: 50 °C, 7 d) before use. All electrolyte preparation was done in a dry atmosphere (dry room: dew point <-65 °C; <5.3 ppm H$_2$O).

#### 2.2.1 Binary liquid electrolytes

For the formulation of IL-based electrolytes, the desired molar fraction of lithium conducting salt was dissolved in the ionic liquid with the same anion, by stirring at 50 °C.

#### 2.2.2 PEO-based ternary polymer electrolytes

For the preparation of the polymer and plasticized polymer electrolytes procedures from the literature[6,25,26] were modified as follows: The lithium salt was mixed with the polyethylene oxide in the desired ratio by manual mixing with a mortar and pestle. Then, (for TSPEs) the IL was added to the blended solids and all components were thoroughly mixed. The mixture was vacuum sealed in a pouch bag and annealed at 80 °C for 72 h, then pressed in a hot-press (80 °C) to the needed thickness (100 µm, 200 µm).

### 2.3 Electrode preparation

LFP cathodes containing TSPEs were prepared following a procedure reported earlier[13]: An electrode paste containing LFP (80.0 wt%), PvdF (7.5 wt%), carbon black (7.5 wt%), PEO-based TSPE (5 wt%, same components and ratio as electrolyte membrane) in NMP was stirred for 24 h at RT. After stirring again at 60 °C for 1 h, the paste was coated on aluminum foil (20 µm) and then dried at 80 °C for 24 h. The electrodes were calendered, punched into 12 mm diameter disks and dried at 80 °C and under reduced pressure (10$^{-3}$ mbar) for 24h. The resulting active mass loading was 1.0 mg cm$^{-2}$.

### 2.4 Differential scanning calorimetry (DSC)

The sample preparation was done under dry conditions and the measurements were done in Tzero™ hermetic aluminum pans on a DSC Q2000 (TA Instruments) calibrated with indium melting point at 156.60 °C. After an isothermal step at 60 °C, the sample were quenched to -150 °C. The heat ramp after the quenching (from – 150 °C to 80 °C; 5 °C min$^{-1}$) was used to characterize the $T_g$s, as after quenching the most distinct $T_g$s are achieved (subsequent cooling ramps (from 80 °C to -150 °C) in the SI, figure S1 and S2).

## 2.5 Pulse field gradient nuclear magnetic resonance

All spectra were recorded on a BRUKER 4.7 T AVANCE III using a diff50 probe. Pulsed field gradient nuclear magnetic resonance (PFG-NMR) data were acquired with a (triply tuned $^7$Li/$^1$H/$^{19}$F) 5 mm coil at 25 °C (±0.2 °C). A 0.25 M LiCl in H$_2$O solution and a 1% H$_2$O in D$_2$O with 0.1% CuSO$_4$ solution ("Doped Water") were utilized for external calibration. The gradient strength was varied from 600 to 2947 G cm$^{-1}$ averaging up to 16 scans with a gradient pulse length $\delta$ of 1 ms and diffusion time $\Delta$ varied from 40 to 200 ms. The self-diffusion coefficients $D$ of the lithium species were derived from a stimulated echo sequence ("diffSte") after fitting the attenuated signal amplitude to the Stejskal-Tanner equation, which describes the case of rather ideal ("free") isotropic diffusion:

$$I = I_0 \times exp\left(-D\gamma^2 \delta^2 g^2 \left(\Delta - \frac{\delta}{3}\right)\right) \quad (1)$$

with $I$ being the signal intensity, $I_0$ the initial signal in the absence of a magnetic field gradient and $\gamma$ the gyromagnetic ratio. Data analysis was done with BRUKER Topspin 3.5 and BRUKER Dynamics Center 2.5.

## 2.6 Electrochemical and physicochemical investigation

### 2.6.1 Viscosity

All viscosity data of binary IL-based electrolytes were acquired with a kinematic Stabinger viscosimeter SVM 3001 from Anton Paar in a temperature range from 10 to 70 °C.

### 2.6.2 Conductivity

The ionic conductivities of the binary IL-based electrolytes were measured using an electrochemical impedance spectroscopy-based conductometer MCS 10 from BioLogic. All conductivity cells with cell constant around 1 were calibrated with a standard 0.1 M KCl solution. The temperature was varied between -20 and 70 °C in steps of 5 °C.

For all ternary polymer electrolytes, the conductivities of the membranes were measured in a coin sell setup between two polished, stainless steel blocking electrodes (Ø = 16 mm). Impedance spectra at 0 – 60 °C were recorded using an Autolab PGSTAT302N potentiostat/galvanostat with impedance spectroscopy function (Metrohm AG). The frequency range was from 1 Hz up to 1 MHz. The thickness of the membrane was controlled before and after the measurement for calculating the cell constant.

### 2.6.3 Li⁺ transference number

Li⁺ transference numbers ($t_{Li^+}$) for the ternary polymer electrolytes were determined electrochemically *via* the Bruce and Vincent method.[27–29] It uses Li|electrolyte|Li symmetric cells and a combination of potentiostatic steps (chronoamperometry) and electrochemical impedance spectroscopy measurement. Symmetric Li|TSPE|Li cells (lithium metal electrodes Ø = 16 mm; TSPE Ø = 18 mm) were assembled for each membrane in a PAT-cell (EL-CELL®). Each cell was rested at 60 °C for 4 days to ensure good contact and stabilized interfaces. Impedance spectroscopy was performed between 100 mHz and 500 kHz prior and at the end of the measurement with an amplitude of 10 mV. For the chronoamperometry a voltage amplitude ($\Delta V$) of 10 mV was applied until the current reached a steady state ($I_{SS}$).

The transference numbers were evaluated *via* equation (2)

$$t_{Li^+} = \frac{I_{SS}(\Delta V - I_0 R_{f,0})}{I_0(\Delta V - I_{SS} R_{f,SS})} \tag{2}$$

with $I_0$ as the initial current, $R_{f,0}$ as initial SEI resistance and $R_{f,SS}$ as SEI resistance in the steady state respectively. Given the uncertainty on the initial current $I_0$, it was calculated *via* impedance spectra collected right before the polarization according to equation (3)

$$I_0 = \frac{\Delta V}{R_{el,0} + R_{f,0}} \tag{3}$$

with $R_{el,0}$ as initial electrolyte resistance.

### 2.6.4 Cycling of LFP||Li metal cells

Galvanostatic cycling experiments with the TSPEs were performed in two-electrode[30] pouch cells (cathode: LFP electrodes containing polymer electrolyte, Ø = 12 mm; anode: lithium metal, Ø = 13 mm) on a Maccor 4000 Battery Tester. After assembly, the cells were rested at open circuit at 60 °C for 24 h. All cycling experiments were then done at 40 °C. The cells were cycled between 2.5 V and 3.8 V *vs.* Li⁺/Li. For the long-term cycling experiments, after 3 formation cycles at C/10, the cells were cycled at C/2 for 300 cycles. For the rate-performance tests the discharge current was increased every 3 cycles (with a constant charge current of C/10) from C/20 to C/10, C/5, C/2, 1C, 2C, then back to C/2 (charge and discharge).

## 2.7 Computational methods

We performed all-atomistic molecular dynamics (MD) simulations of two polymer electrolyte mixtures using the software package GROMACS (version 2018.8).[31–34] Both systems comprise 10 coiled PEO chains, which each contain 54 ether oxygen (EO) units, as well as 54 lithium salt and 54 ionic liquid ion pairs. This aims to reproduce the experimentally investigated membrane composition PEO:salt:IL of 20:2:2. The lithium salt and $Pyr_{14}^+$-based IL share the same anion, for which either TFSI or its asymmetric analog TFSAM is employed.

The atomic interactions of PEO were parameterized by the optimized potentials for liquid simulations all-atom (OPLS-AA) force field (FF)[35], while the interactions of the ionic constituents, *i.e.*, $Li^+$, $Pyr_{14}^+$ and the anions TFSI and TFSAM, were modeled by the widely recognized OPLS-AA-derived *CL&P* force field.[36–40]

Transport properties are commonly reported to be underestimated when employing non-polarizable force fields, however, consideration of polarization effects comes at a much greater computational cost. In order to mimic an effective charge screening in a mean-field like manner, the atomic partial charges were scaled by a uniform factor of 0.8.[41–45]

The initial structures were generated using PACKMOL,[46] which randomly distributed the molecules in a cubic cell in the gas phase. Then, the systems were relaxed by means of an equilibration scheme: After an energy minimization the systems were pre-equilibrated at a temperature of 500 K and pressure of 1 bar for 10 ns with a time step of 0.5 fs in the *NpT* ensemble, where the temperature was maintained by a velocity-rescale thermostat and the pressure by a Berendsen barostat.[47,48] Ensuing another energy minimization, the systems were cooled to 400 K and further equilibrated for 300 ns employing an increased time step of 2 fs. Prior to the production run the systems are further propagated for 40 ns using a Parrinello-Rahman barostat to control the pressure.[49] The subsequent production runs were carried out in the *NpT* ensemble at 400 K and a pressure of 1 bar by means of a v-rescale thermostat ($\tau_T$ = 1 ps) and a Parrinello-Rahman barostat (compressibility of $4.5 \cdot 10^{-5}\ bar^{-1}$, $\tau_p$ = 5 ps). The equations of motion were integrated using the leap-frog algorithm at a time step of 2 fs and the coordinates were saved every 2 ps. The produced trajectories have a total length 2 $\mu$s. To prevent the system from drifting, *i.e.*, the accumulation of a center-of-mass (com) translational velocity, the com motion was removed at every step.

The smooth Particle-Mesh Ewald method was used to compute electrostatic interactions,[50] relying on a grid spacing of 1 Å, as well as an interpolation-order of 6. The cut-off distances for long range electrostatic and the van der Waals interactions were both set to 14 Å, and the hydrogen bonds were constrained using the linear constraint solver (LINCS).[51,52]

The simulations were analyzed using the GROMACS toolkit[53] as well as customized scripts relying on the Python library MDAnalysis.[54,55]

# 3  Results

## 3.1  Fundamental understanding of the unique Li–TFSAM interaction in ILs

As mentioned above, the TFSAM anion is utilized to enhance the Li$^+$ mobility in TSPEs *via* a novel 'anion-assisted' transport mechanism. Nevertheless, to fully understand this new transport mode, it is necessary to take a step back and investigate the interaction of Li$^+$ with asymmetric anions like TFSAM in a more simplified system (than in the TSPE). Here, the system of choice is the binary mixture of an IL and the corresponding lithium salt to gain a fundamental understanding first.

Regarding the Li$^+$ coordination by anions with different functional groups, it has been reported that in IL mixtures including TFSI and DCA, the CN-group of DCA is the preferred coordination site.[41] Recently, Nürnberg et al. reported that, at concentrations of LiTFSAM below 30 mol%, only Li–CN coordination is observable *via* Raman and NMR spectroscopy and that this Li–anion interaction is stronger compared to that of Li–TFSI.[56] A recent computational counterpart study confirmed that only in the regime of high salt concentrations when coordination can no longer be afforded by the preferred cyano-group, new binding geometries, *e.g.*, to the sulfonyl oxygens or even to the center nitrogen, emerge.[57]

To complete the picture, here, TFSAM containing electrolytes are compared to both of its structurally related symmetric anions (TFSI and DCA, see Figure 1). To precisely observe any effects on Li–anion interaction, a comparative study with IL-based electrolytes with incremental increases of lithium salt fractions (0 – 12.5 mol%) was performed. In the following combined physicochemical and electrochemical study, the unique Li–TFSAM interaction is highlighted in IL-based electrolytes, which grants a proper background to understand the change in of Li$^+$ transport, when TFSAM is utilized in TSPEs later.

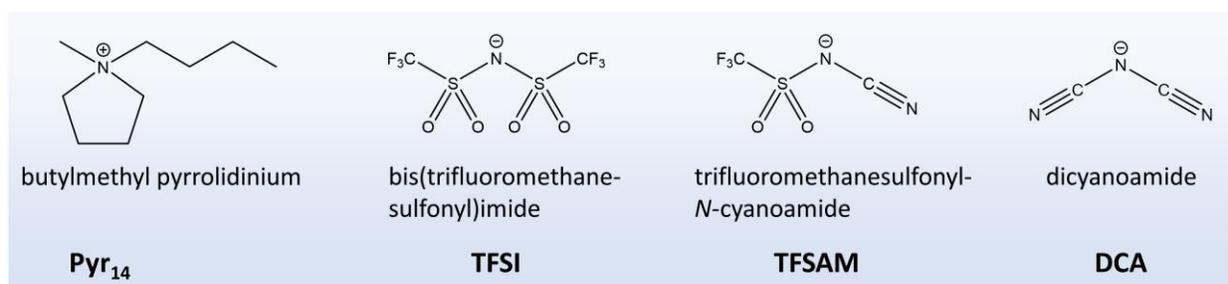

Figure 1: Molecular structures of butylmethyl pyrrolidinium (Pyr$_{14}$), bis(trifluoromethanesulfonyl)imide (TFSI), trifluoromethanesulfonyl-*N*-cyanoamide (TFSAM) and dicyanoamide (DCA).

### 3.1.1 TFSAMs effect on viscosity, ionic conductivity and Li$^+$ diffusion

As reported earlier,[23] the viscosities of the pure ILs, Pyr$_{14}$DCA, Pyr$_{14}$TFSAM and Pyr$_{14}$TFSI, scale with the size of the anion, *i.e.,* with the larger anions leading to higher viscosities, as in ILs the molecular radius of the ions usually affects the shear resistance (and therefore the viscosity) to a much greater extent than the interactions between the Pyr$_{14}^+$ cation and the anion.[58–60] This behavior changes when lithium ions are introduced in the system. As can be seen in Figure 2a, the introduction of lithium salt has a higher influence on the viscosities of TFSAM-based binary electrolytes, than on either TFSI or DCA-based ones. At low salt contents, the viscosity curves still follow the previous trend, *i.e.,* grow according to anion size. This changes with an increasing fraction of Li$^+$ and, at 12.5 mol%, the TFSAM-based electrolyte even surpasses its TFSI counterpart. High interactions between Li$^+$ and TFSAM$^-$, leading to increased shear resistance, are a possible explanation, which is further corroborated by the quite different coordination shell sizes observed in MD simulations.[57] Although, on average, three TFSI molecules closely entwine themselves around the lithium ion in a preferably bidentate binding geometry, four TFSAM anions create an extended solvation sphere through solely mono-dentate cyano-contacts.

A similar trend is observed for the ionic conductivity of these binary electrolytes (Figure 2b). A drop of conductivity upon adding a lithium salt is typical for IL-based electrolytes,[61–63] as the Li–anion interaction is stronger than the interaction between the anion and Pyr$_{14}^+$ (sterically and electronically hindered: +*I* effect of the alkyl chains). Here also, the Li$^+$ coordination by TFSAM seems particularly high. It has already been reported that, for TFSAM, the decrease of conductivity with increasing salt concentration is more severe (compared to TFSI)[56] due to the higher ion–ion interaction because of the Li–CN coordination in the case of TFSAM. Thus, one could have expected that within the series of TFSI, TFSAM, DCA, the DCA-based electrolytes would have shown the highest conductivity drop with salt addition, given the presence of two -CN groups on the anion. However, this is not the case. As is visualized more clearly in Figure 2c, the TFSAM-based electrolytes show the highest relative drop in conductivity compared to the corresponding pure IL.

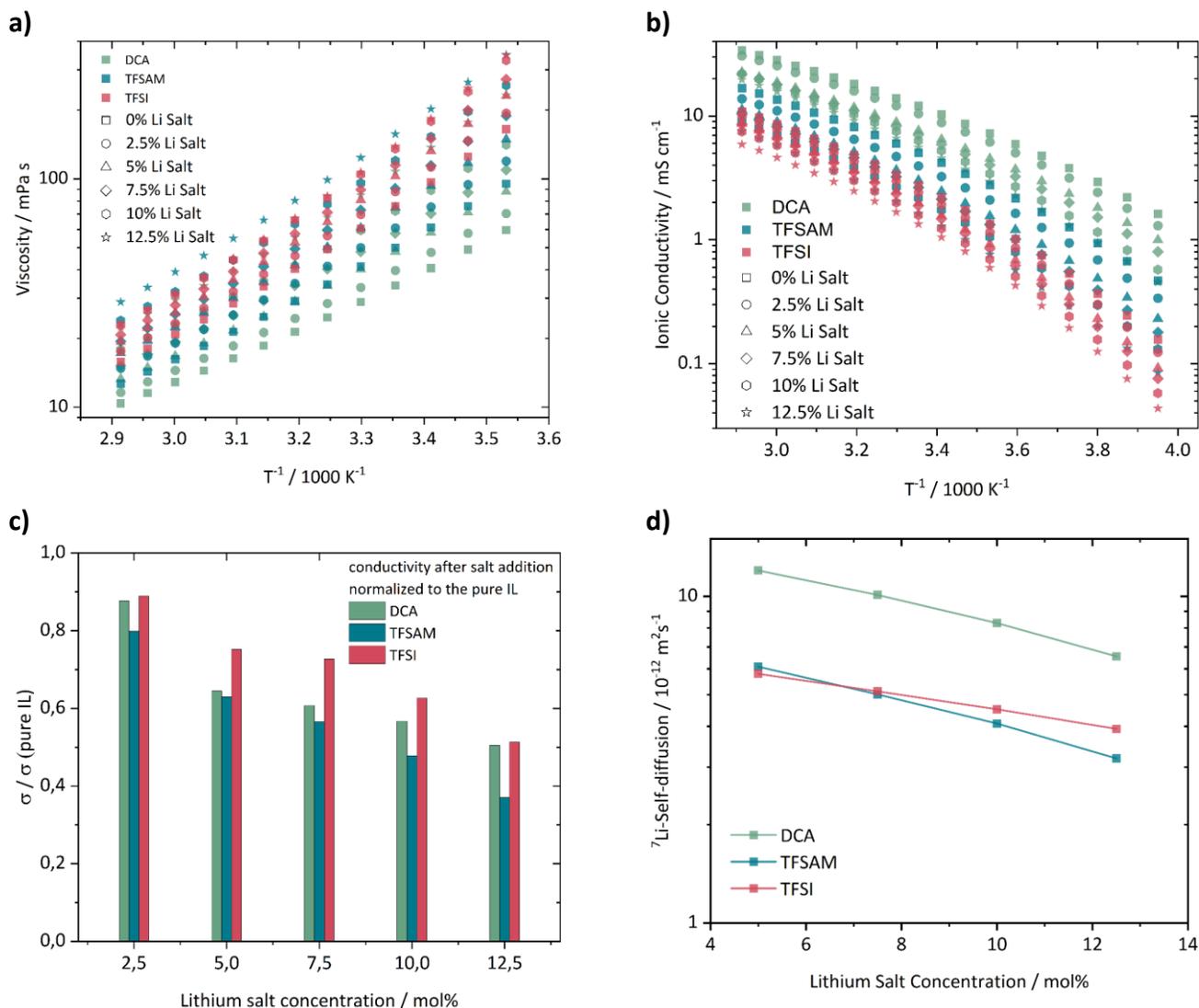

Figure 2: **a)** Viscosities of the binary IL-based electrolytes depending on temperature (10 – 70 °C) and lithium salt concentration (0 – 12.5 mol%); **b)** ionic conductivities of the binary IL-based electrolyte depending on temperature (-20 – 70 °C) and lithium salt concentration (0 – 12.5 mol%); **c)** drop of conductivity with addition of lithium salt; ionic conductivities of binary IL-based electrolytes divided by the conductivity of the pure IL; exemplarily shown for 40 °C; **d)** concentration dependent Li$^+$ diffusion coefficients determined *via* PFG-NMR at 25 °C.

This behavior is logically also reflected in the Li$^+$ diffusion coefficients, which were determined by PFG-NMR for these binary electrolytes (Figure 2d). Interestingly, the TFSAM-based electrolytes reach values close to TFSI-based ones at 7.5 mol% salt content, although at this concentration the TFSAM electrolyte has a higher conductivity and a lower viscosity. Thus, a hindered Li$^+$ transport due to its strong coordination by TFSAM can be assumed for these binary mixtures. The introduction of the stronger coordinating CN-group seems to have a more pronounced effect in the case of TFSAM compared to DCA. When explaining this behavior, the obvious difference is that DCA offers two equally favored coordination sites for Li$^+$, which might be counter-intuitive at first. Nevertheless, as the electron density of the negative charge can be delocalized over both CN-groups of DCA (see SI, figure S3, for mesomeric structures), the Li–DCA interaction of an existing ion pair would be weakened as soon as another $Li^+$ approaches the opposite CN-group, leading to an overall faster ion exchange. On the other hand, with TFSAM, this is less likely to happen because the CN-coordination site of TFSAM is electronically much favored over the $Li^+$-interaction with an oxygen of the $SO_2CF_3$-group. Furthermore, the size difference might play a role with the small DCA anion being faster by itself and therefore moving faster from the solvation sphere. Thus, it can be stated that the asymmetric combination of electron-withdrawing groups creates in the case of TFSAM an "anomaly" within the series of TFSI, TFSAM and DCA in terms of Li–anion interaction.

To conclude, the investigation of the binary systems supports the expected strong Li–anion interaction for TFSAM with a preferred Li–CN coordination. Additionally, the direct comparison with both related symmetric anions (TFSI and DCA) also reveals that it is not only because the Li–CN coordination is "naturally" stronger than the Li–O. Instead, the asymmetric combination with a weaker coordination side (-$SO_2CF_3$) lead to an even more pronounced Li–TFSAM interaction on the CN-site. In the case of binary liquid electrolytes, this has a negative effect on $Li^+$ transport in view of battery application.

However, in the following, this marked coordination is utilized to shift the $Li^+$ coordination and transport in PEO-based electrolytes from polymer to anion dominance and the binary (IL-based) systems results grants support for the understanding of the transport mechanism in TSPEs.

### 3.2 TFSAM in ternary solid polymer electrolytes

TFSAM with its strong unilateral $Li^+$ coordination was selected for a new class of TSPEs, where the IL does not act only as plasticizer, but at the same time, can 'free' the $Li^+$ from the strong coordinating PEO chain and thereby accelerate $Li^+$ transport. Thus, TFSAM-based TSPE membranes with several PEO:salt:IL ratios were prepared by a solvent-free processing and investigated. They are compared to state-of-the-art TFSI-based TSPEs in terms of transport properties and, with the support of MD simulations, a new 'anion-assisted' transport mechanism is proposed (DCA-based TSPEs could not be prepared due to fast occurring phase separation of IL and PEO).

#### 3.2.1 Thermal behavior and $Li^+$ coordination environment

The thermal behavior of TFSI- and TFSAM-based polymer electrolytes was investigated *via* differential scanning calorimetry (DSC). Thus, membranes with a favorable PEO:salt:IL ratio (in terms of crystallinity) can be preselected. At the same time, the glass transition temperatures ($T_g$) of the membranes grant first insights on potential difference on a microscopic level.

The DSC heating scans of the TSPEs are presented in Figure 3a (heating scan after a quenching step shown; following cooling scan: see SI). There are only minor differences in the crystallization and melting behavior between TFSI- and TFSAM-based membranes, though it is noticeable, that TFSI-based membranes tend to stay more amorphous as seen from the smaller or non-existent cold-crystallization peaks. This is to be expected, as TFSI is the lower coordinating anion it can be considered to be fully dissociated in the PEO-matrix, whereas the TFSAM anion although asymmetric, likely suffers from stronger interaction with Li$^+$ that help the formation of crystalline complexes from the stronger solvation structures in solution. In general, for both anions, a PEO:salt ratio of 20:2 seems to be beneficial to reduce the crystallinity.

a) 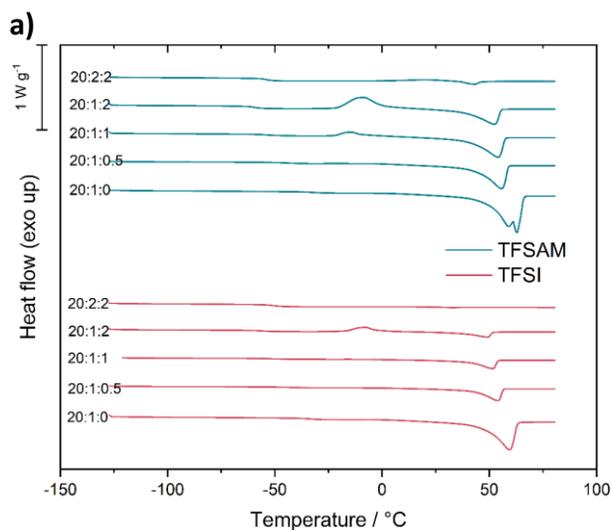

b) 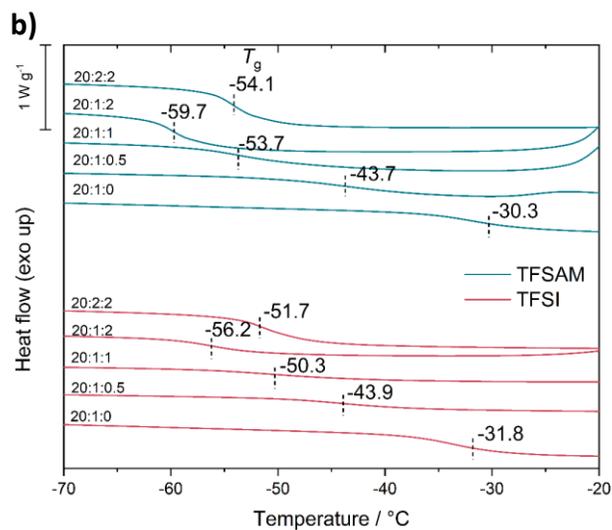

c) 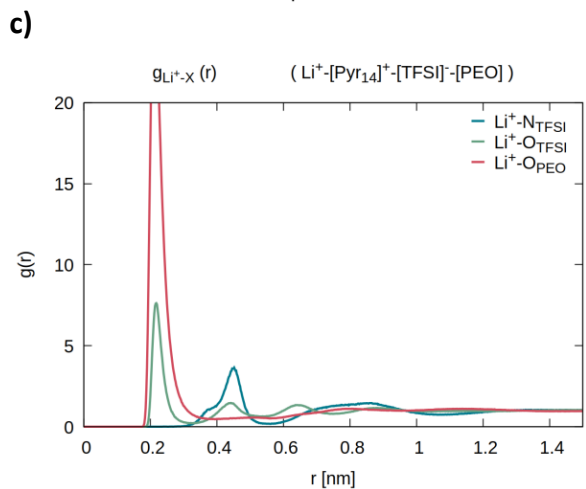

d) 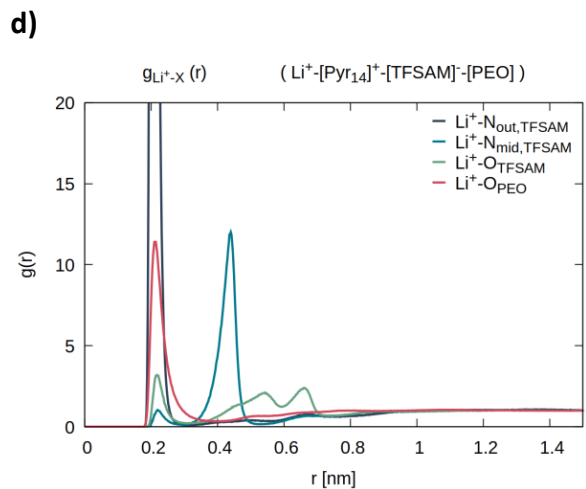

e)

f)

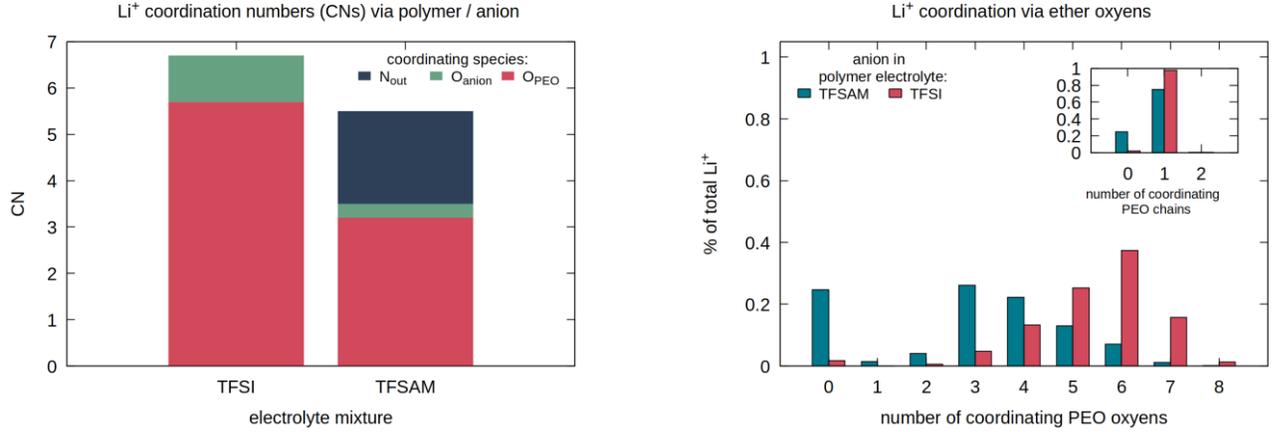

Figure 3: **Top row a) and b)**: DSC thermograms of TFSAM and TFSI based ternary electrolytes with different PEO:salt:IL ratios; scan rate: 5 °C min$^{-1}$; shown: heat ramp after quenching step; **middle row**: Coordination environment of the lithium ion displayed *via* radial distribution functions $g_{Li^+-X}(r)$ between lithium ions and binding sites provided by the polymer chains and the respective anion, *i.e.*, TFSI **c)** and TFSAM **d)**; **e)** composition of lithium coordination shell in terms of ether oxygen and anion coordination numbers CN; **f)** Probability distribution of $Li^+ - O_{PEO}$ coordination numbers regardless if the ether oxygens are provided by a single or multiple PEO chains. The inset shows the probability to find Li$^+$ bound to one or two PEO chains or structurally decoupled from the polymer (0).

Besides crystallinity, the evolution of the $T_g$ is of high interest for polymer electrolytes, as a lower $T_g$ usually represents a higher segmental mobility of the polymer chains at a given temperature, hence, a faster Li$^+$ transport. With the first addition of lithium salt (see 20:2:0 membranes in Figure 3b), the usual increase in $T_g$ (compared to pure PEO: -52 °C[64]) is observed, as the Li$^+$ coordination by the ether oxygens of the PEO 'stiffens' the polymer chains. After that, the addition of IL seems to have the usual plasticizing effect of increasing the segmental mobility again (seen as a lowering the $T_g$). Interestingly, the use of Pyr$_{14}$TFSAM leads to lower $T_g$ than Pyr$_{14}$TFSI. One could argue that Pyr$_{14}$TFSAM is the better plasticizer, but this would be in conflict with the results in binary electrolytes above, where TFSAM-based electrolytes show even higher viscosities (at high salt ratios) than TFSI-based. The more likely explanation is that, due to the introduction of the coordinating TFSAM anion, the coordination of Li$^+$ by PEO (and its influence on the $T_g$) is affected.

The molecular resolution of the MD simulations provides the opportunity to elucidate the lithium coordination behavior in the TFSI and TFSAM-containing membranes for qualitative differences. The solvation environment of the lithium ions is analyzed by means of radial distribution functions (RDFs) $g_{Li^+-X}(r)$ between lithium and the possible binding sites $X$ offered by the polymer and the anions (equation (4)):

$$g_{Li^+-X}(r) = \frac{V}{4\pi r^2 N_{Li^+} N_X} \langle \sum_i^{N_{Li^+}} \sum_j^{N_X} \delta(r - |\vec{r}_i - \vec{r}_j|) \rangle \quad (4)$$

where $V$ corresponds to the volume of the simulation cell, and $N_{Li^+}$ and $N_X$ denote the number of atoms of the respective species. In principle, the RDF probes the probability to encounter species $X$ within a distance $r$ of a distinct Li$^+$. Figure 3c and d depict the arrangement of the polymer ether oxygens $O_{PEO}$ and the anions around lithium. To probe both the number of coordination contacts and the binding geometry, *i.e.*, monodentate *vs.* bidentate, the TFSI coordination is tracked by its $O_{TFSI^-}$ atoms as well as $N_{TFSI^-}$. As previously discussed, TFSAM provides a cyano-nitrogen as an additional, and most favorable, coordination site which is termed $N_{out,\ TFSAM^-}$ opposed to the central nitrogen $N_{mid,\ TFSAM^-}$ which is also contained in TFSI.

In both electrolyte mixtures, we observe sharp coordination peaks at a distance of 2 Å which therefore describe the compositions of the first coordination sphere. In the TFSI-containing electrolyte, lithium is primarily solvated by the ether oxygens but also shows contributions from $O_{TFSI^-}$. Unlike in the pure IL scenario, the latter binds to lithium in a monodentate manner which can be deduced from the split peak structure of $Li^+$-$N_{mid,\ TFSAM^-}$ RDF with the right shoulder being more populated.

In accordance with previous experimental and theoretical studies involving cyano-moieties in competition with other Li$^+$ coordination sites such as $O_{PEO}$ or $O_{TFSI^-}$,[41,56,57,65] we find that the $N_{out,\ TFSAM^-}$ coordination peak superimposes that of the ether oxygens, with the latter being downsized significantly in reference to the TFSI analog mixture. We can thus infer that lithium coordination by the polymer chain is partially superseded by TFSAM.

The coordination numbers (CNs) are computed by integrating $g_{Li^+-X}(r)$ up to the position of the first minimum $r_{min}$ (equation (5)):

$$\text{CN} = 4\pi\rho_{X,\text{bulk}} \int_0^{r_{min}} dr\ r^2 g_{Li^+-X}(r) . \tag{5}$$

The average composition of a lithium solvation shell in both electrolytes is compared in Figure 3e and visualized in snapshots in Figure 4.

In good agreement with a recent simulation study employing a polarizable force field of a similar system composition (10:1:2),[66] we observe for the TFSI-containing mixture a distribution of coordination motifs involving 4-7 ether oxygens, which are almost exclusively provided by a single PEO-strand. Deviating from this simulation study, we find a statistically more frequent additional coordination *via* one TFSI oxygen on average. Despite scaling the atomic partial charges, the strength of ionic interactions may still be slightly overestimated and the cause for such discrepancies. For the polymer electrolyte comprising the TFSAM anion on the other hand, the coordination shell is considerably downsized. First, the absolute number of atomic species that are bound to lithium reduces from 6.7 in the TFSI analog to 5.5 (see Figure 3e) of which only about 3 monomer units account for the structural attachment of lithium to the polymer chains. Instead, the lithium coordination environment reveals an increasing anionic proportion *via* 2 cyano-moieties $N_{out,\ TFSAM^-}$ and a small contribution from 0.3 TFSAM oxygens.

When examining the underlying distribution of ether oxygen coordination numbers $CN_{O_{PEO}}$, we see that not only is the crown-ether like wrapping of lithium partly suppressed by the presence of the anion, but also a quarter of the lithium ions is entirely liberated from the polymer host (see inset in Figure 3f).

To this end, the coordination analysis indicates that the strongly coordinating TFSAM anion induces a decoupling of lithium from the polymer which, in turn, might contribute to a higher chain flexibility and could rationalize the lower glass transition temperatures observed for the TFSAM containing membranes.

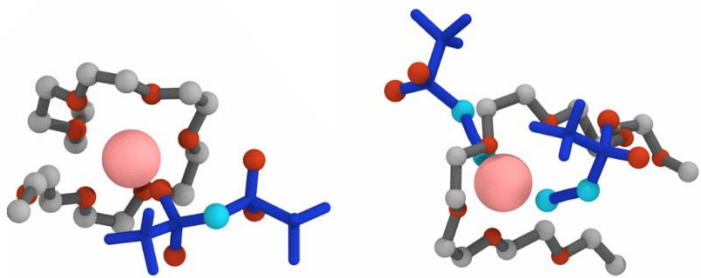

Figure 4: Snapshots depicting the average lithium coordination environment in the respective electrolyte mixture, *i.e.*, the TFSI-based (left) and TFSAM-based (right) polymer electrolyte. For reasons of clarity only the coordinating section of the polymer chain is displayed. Oxygen atoms are shown in red and nitrogen atoms in cyan.

### 3.2.2 Transport properties: ionic conductivity, Li⁺ transference numbers, Li⁺ conductivity

Since TFSAM changes the coordination sphere of Li⁺ in PEO-based TSPEs, it is worth investigating how this influences the dynamic processes. With regard to the application of TSPEs in lithium batteries, the Li⁺ transport properties as well as the actual transport mechanism are of high interest:

The ionic conductivities of membranes with PEO:salt:IL ratios of 20:2:1 and 20:2:2 are shown in Figure 5a. As usually observed, a higher fraction of IL in the membrane leads to an increase in conductivity. Moreover, it can be seen that TFSAM-based electrolytes have lower conductivities than TFSI analogs, for instance at 40 °C: 6.1 mS cm$^{-1}$ for TFSI 20:2:2, *vs.* 3.7 mS cm$^{-1}$ for TFSAM 20:2:2 and 2.9 mS cm$^{-1}$ for TFSI 20:2:1 *vs.* 1.3 mS cm$^{-1}$ for TFSAM 20:2:1. However, considering the lithium battery application, the mobility of Li⁺ is a more crucial factor than the total ionic conductivity of the electrolyte.

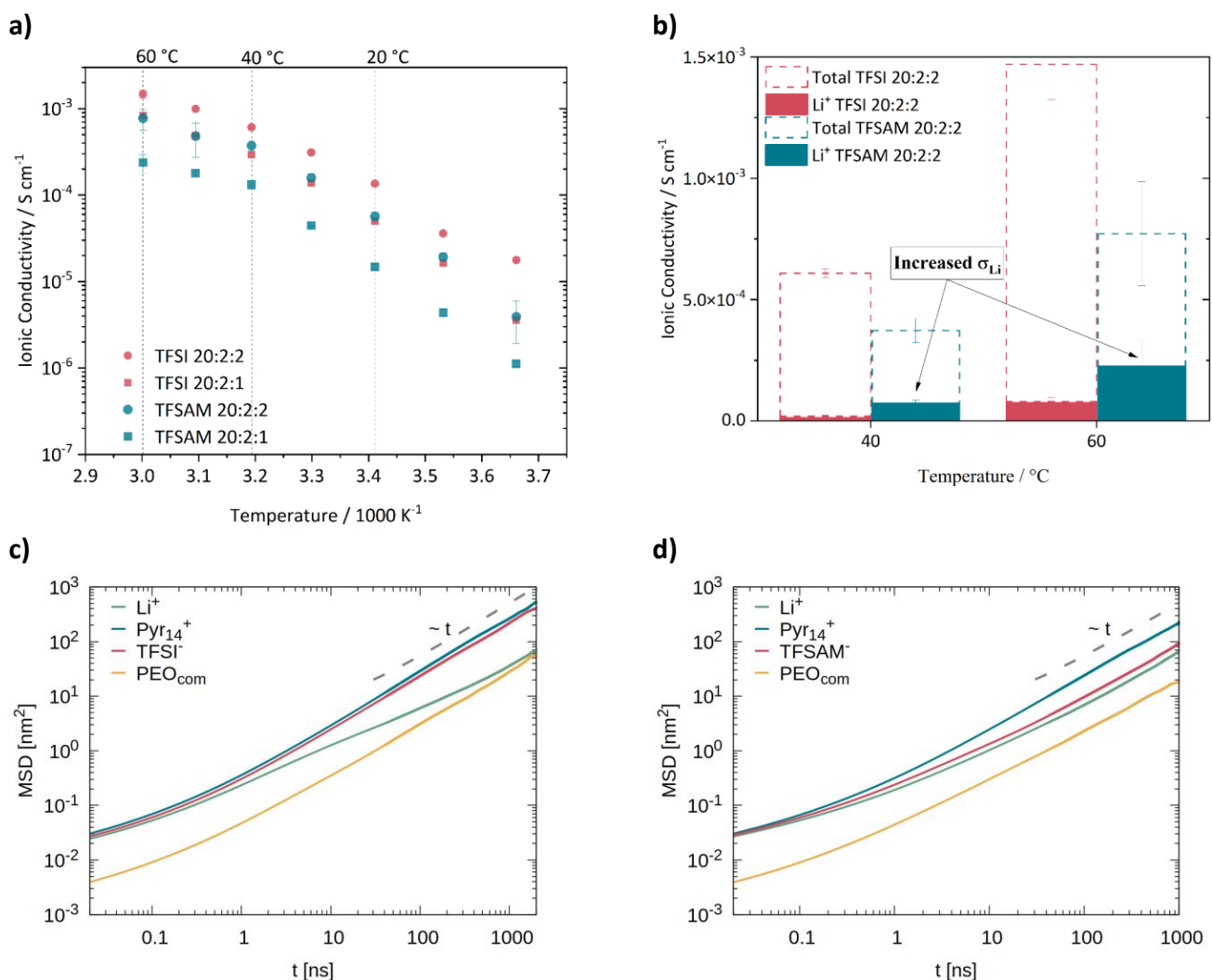

Figure 5: **Top row: a)** Temperature dependent ionic conductivities of TFSAM and TFSI based ternary electrolytes with PEO:salt:IL ratios 20:2:1 and 20:2:2; **b)** total ionic conductivity and portion of the conductivity contributed by Li⁺ at 40 and 60 °C; **bottom row**: mean squared displacements of all molecular constituents in **c)** the TFSI-based and **d)** the TFSAM-based mixture.

Therefore, the Li⁺ transference numbers ($t_{Li^+}$) of the membranes were determined *via* the Bruce and Vincent method.[27–29] As displayed in Table 1, $t_{Li^+}$ is increased drastically when switching the anion from TFSI to TFSAM, *e.g.* at 40 °C a more than 6-fold increase from 0.03 to 0.20 (the values for TFSI are coherent with results derived from both electrochemical and PFG-NMR measurements on a TFSI 20:2:2 membrane ($t_{Li^+}$ = ca. 0.05), with half the IL content[13]).

Table 1: Electrochemically measured Li⁺ transference numbers of TFSAM and TFSI based TSPEs with a PEO:salt:IL ratio of 20:2:2.

| Temperature / °C | Li⁺ Transference Number | | |
|---|---|---|---|
| | TFSI 20:2:2 | TFSAM 20:2:2 | Increase |
| 40 | 0.03 ± 0.01 | 0.20 ± 0.01 | 667% |
| 60 | 0.05 ± 0.01 | 0.29 ± 0.06 | 600% |

The transference numbers were used to calculate the fraction of the ionic conductivity corresponding to lithium ions ($\sigma_{Li^+}$) *via* equation (6):

$$\sigma_{Li^+} = \sigma_{total} \cdot t_{Li^+} \tag{6}$$

with $\sigma_{total}$ being the total ionic conductivity of the TSPEs. Figure 5b illustrates how the change from TFSI to TFSAM improves $\sigma_{Li^+}$ in the membrane, although the total ionic conductivity is lower. As mentioned earlier, in the case of TFSI, the effects of the introduction of the IL to PEO-based polymer electrolytes is well-known: The IL acts mostly as plasticizer, making the polymer segments more mobile and thus increasing conductivity and Li⁺ mobility,[18,67] while direct Li–TFSI interactions are very limited within these TSPEs.[26] Therefore, with TFSI-based ILs as plasticizers, the Li⁺ transport is increased (compared to PEO–salt 'dry' polymer electrolytes) but the usual transport mechanisms for Li⁺ remain the same. Keeping this in mind, it becomes clear that the introduction of TFSAM-based ILs affects the conductivity and the Li⁺ transport in TSPEs in a completely different way that results in the drastic increase of $t_{Li^+}$ and $\sigma_{Li^+}$. So, it is safe to assume that the unilateral Li–anion coordination of TFSAM, changes the whole transport mechanism for Li⁺ in PEO-based TSPEs, and that this change is beneficial and leads to enhanced Li⁺ mobility.

The transport characteristics of all species are probed by the mean-squared displacements (MSDs) with a larger MSD implying faster dynamics (equation (7)):

$$\mathrm{MSD}_i(t) = \langle (\vec{r}_i(t+t_0) - \vec{r}_i(t_0))^2 \rangle \tag{7}$$

where $\langle ... \rangle$ denotes the ensemble average over all particles of species $i$ and possible starting times $t_0$. Due to the highly viscous nature of the electrolyte mixtures, the simulation time required to reach the diffusive regime, *i.e.*, $\mathrm{MSD}_i(t) \propto t$, is considerably long. Since MSDs are related to the self-diffusion coefficients $D_i$ through the Einstein relation $D_i = \lim_{t \to \infty} \mathrm{MSD}_i(t) / 6t$, the qualitative ranking of the diffusivities can be deduced from the slopes of the respective $\text{MSD}_i$ at long times.

Figure 5c and d displays the $\text{MSD}_i(t)$ of the center-of-mass (com) of all species in both electrolytes. In the TFSI-containing electrolyte, the diffusivities rank as $D_{\text{Pyr}_{14}^+} > D_{\text{TFSI}^-} > D_{\text{Li}^+} \gtrsim D_{\text{PEO}_{\text{com}}}$. It is commonly found in the polymer electrolyte literature that lithium dynamics are strongly coupled to that of the polymer segments[18,66–68], which is reflected in the approach of $\text{MSD}_{\text{Li}^+}$ and $\text{MSD}_{\text{PEO}_{\text{com}}}$ at long times. Note, however, that for very long times, the Li$^+$ dynamics exceeds that of PEO due to Li$^+$ transfer between distinct PEO chains.

Interestingly, a different picture emerges with the TFSAM anion for which the ranking is qualitatively maintained $D_{\text{Pyr}_{14}^+} > D_{\text{TFSAM}^-} \gtrsim D_{\text{Li}^+} > D_{\text{PEO}_{\text{com}}}$. However, the lithium mobility shifted up towards that of TFSAM and is therefore substantially enhanced in comparison to the polymer chains. In the context of the structurally indicated lithium-polymer uncoupling, the similar lithium and TFSAM diffusivities suggests their collective motion. This assumption is further corroborated by the comparison of mean binding times $\tau_{\text{Li}^+-X}$ of $\text{Li}^+$ to either a distinct anion or polymer chain. As shown in Table 2, the introduction of the strongly coordinating TFSAM anion results in a dramatic shift of anion *vs.* polymer host-related time scales: whereas the time a distinct $\text{Li}^+$ spends on average in the neighborhood of the same anion is increased by a factor 10 in the TFSAM electrolyte, the binding time to a distinct PEO chain drops to less than 10% compared to the TFSI system. This suggests that the long-range lithium transport in the TFSAM-based electrolyte is substantially aided by frequent lithium inter-chain transfers.

Table 2: Mean residence times $\tau_{Li^+-X}$ of Li$^+$ with a specific TFSI or TFSAM molecule or a distinct PEO chain. $\tau_{Li^+-X}$ is evaluated from the respective residence time autocorrelation function (ACF) which probes the probability $p_{Li^+-X}(t)$ that a Li$^+$- $X$ pair is preserved after time $t$. The ACF is fitted by a stretched-exponential decay whose integral yields an estimate of the average binding time.[57]

|   | $\tau_{Li^+-anion^-}$ | $\tau_{Li^+-chain}$ |
|---|---|---|
| **TFSI system** | 3.2 ns | 1171.9 ns |
| **TFSAM system** | 32.1 ns | 75.6 ns |

With these additional insights, a novel transport mechanism can be proposed: Figure 6 displays the discovered change to the 'anion-assisted' Li$^+$ transport when using a coordinating anion like TFSAM. As described above, the introduction of TFSAM loosens the PEO–Li coordination (observable in the shorter mean resistance times, the changed coordination sphere, the higher Li$^+$ mobility (MSD)). As displayed in the scheme, Li$^+$ inter-chain transport is now enhanced (deduced from the mean residence times and MSDs). Moreover, this 'anion-assisted' Li$^+$ transport is faster than in a TFSI-based TSPE (shown in the $t_{Li^+}$ and $\sigma_{Li^+}$ and the MSDs).

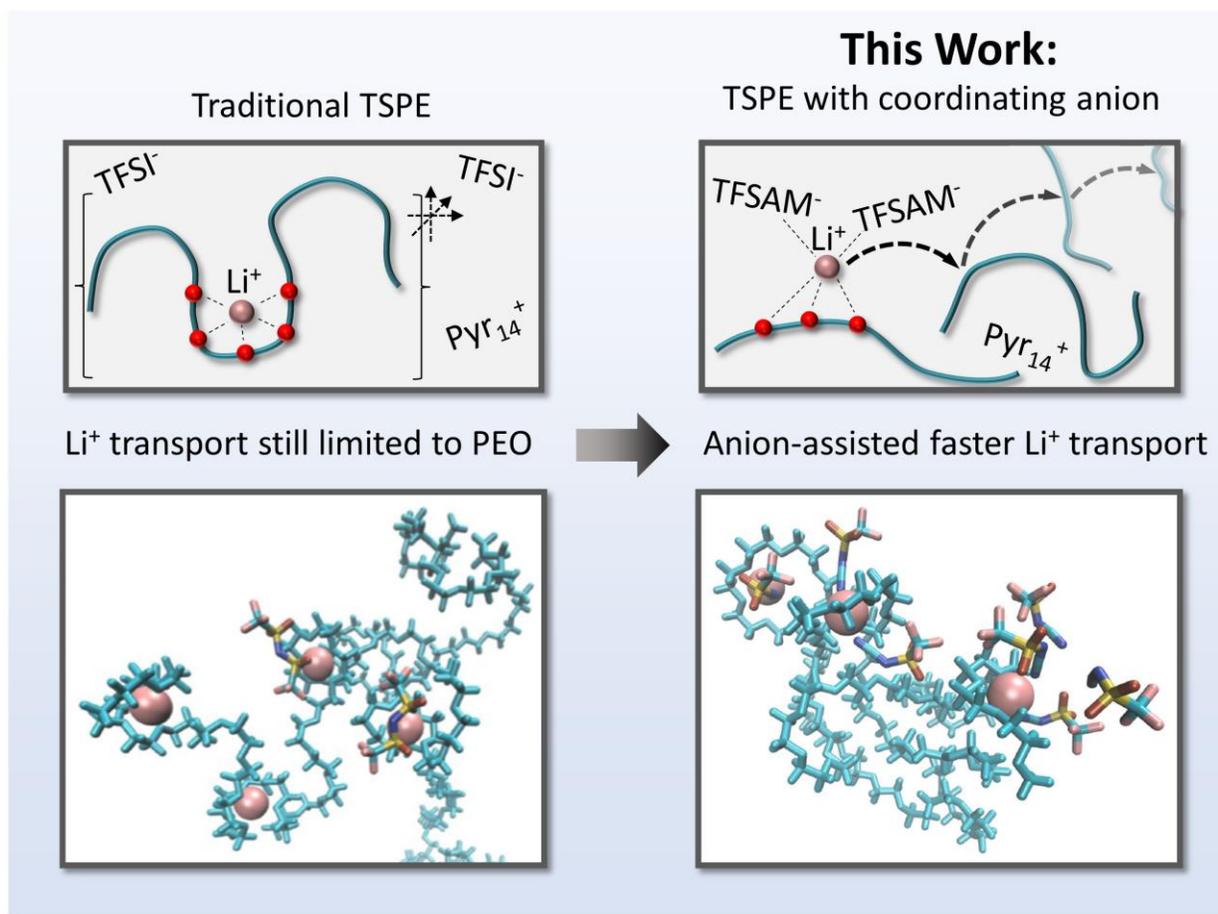

Figure 6: Novel, faster 'anion-assisted' transport mechanism (TFSAM-based) compared to the traditional Li$^+$ transport in TFSI-based TSPEs.

## 3.3 TFSAM-based TSPEs in lithium metal polymer cells

The applicability of TSPEs incorporating TFSAM was tested in LMPBs: As one of the advantages of PEO-based electrolytes is the ability to form an effective stable solid electrolyte interphase on lithium metal,[69] this is the obvious anode material of choice for this study, also with regard to the recent trend to "revive" lithium metal as next generation anode.[70,71] When it comes to cathode materials, it seems that in practical applications, the electrochemical stability of PEO is still the limiting factor.[69,72] Therefore, this proof of concept was carried out in LFP‖Li metal cells and compared to TFSI-based LFP‖Li cells tested in parallel. Figure 7a shows the specific capacity evolution and Coulomb efficiency of the cells over 300 cycles.

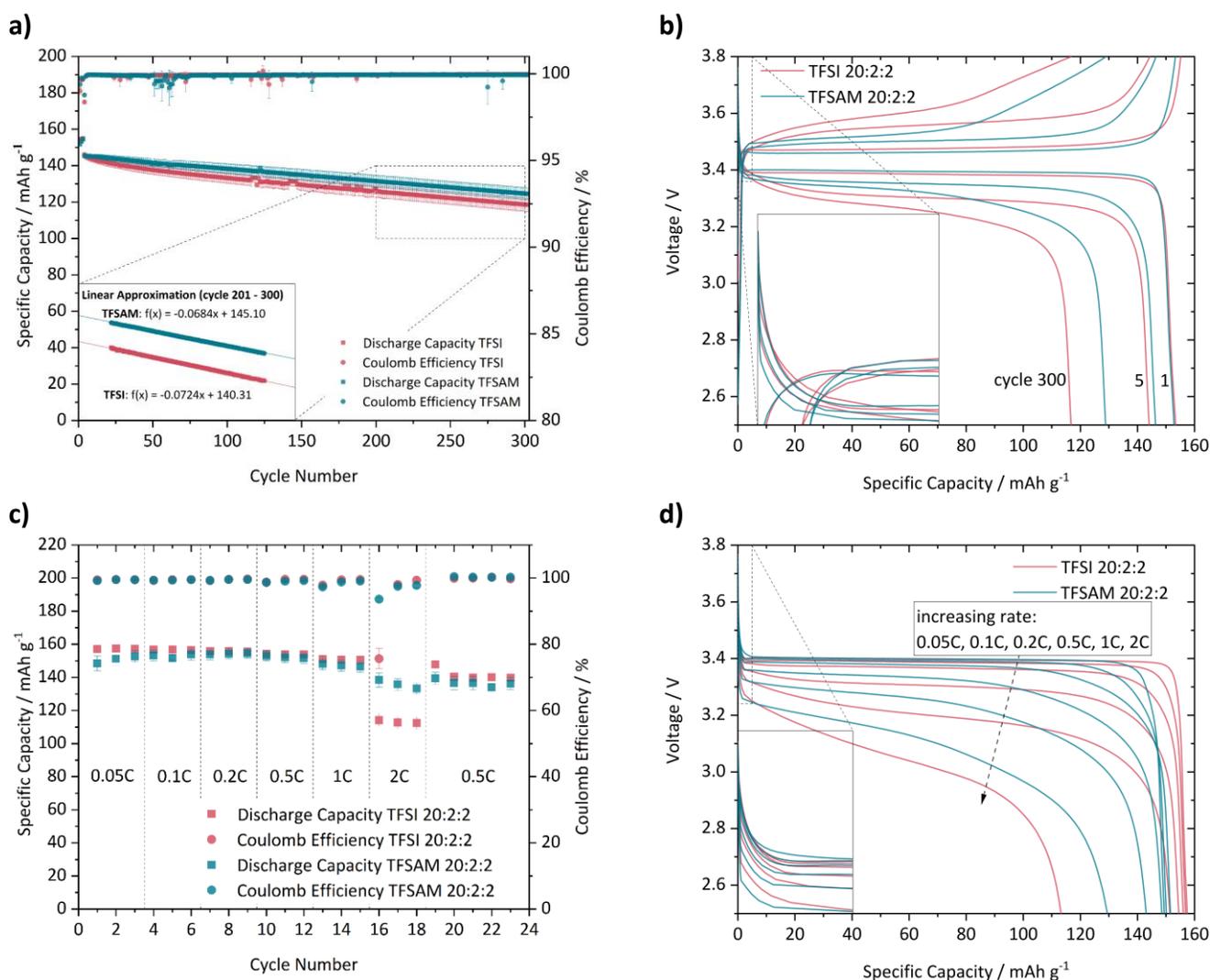

Figure 7: **a)** Galvanostatic cycling (300 cycles) of LFP‖Li metal cells with TFSI and TFSAM based TSPEs; 3 formation cycles: C/10, then C/2; mass loading: 1.1 mg cm$^{-2}$; 3 cells per material for error calculation; **b)** voltage profiles of selected charge and discharge curves of the long-term cycling (300 cycles); **c)** galvanostatic cycling (rate-performance test) of LFP‖Li metal cells with TFSI and TFSAM based TSPEs; charge rate: C/10, increasing discharge rate: C/20 to C/10, C/5, C/2, 1C, 2C, then back to C/2 (charge and discharge); mass loading: 1.1 mg cm$^{-2}$; **d)** voltage profiles of selected discharge curves of the rate-performance test, 3$^{rd}$ cycle of each C-rate shown.

Cells with TFSAM-based TSPEs show, on average, a slightly higher first cycle efficiency of 99.4%, whereas those with TFSI-based TSPEs are around 99.0%, which is close but does not reveal any particular instability of TFSAM *vs.* TFSI. Over the course of 300 cycles at C/2, TFSAM demonstrates an advantage in terms of capacity retention with 86% of the initial capacity (at cycle 300, compared to the 5$^{th}$ cycle), over TFSI with an 82% capacity retention. It can be mentioned that, especially within the first 50 cycles, the capacity decay differs between the two systems, with the TFSI cells losing capacity faster. However, upon further cycling, the cells with TFSI seem to stabilize and adopt a similar capacity fading rate. In the last 100 cycles, the capacity decay is well described by a linear fit for both systems (determination coefficients: $R^2$(TFSI) = 0.9998, $R^2$(TFSAM) = 0.9995). From the steepness of this linear development (TFSI: -0.072 mAh g$^{-1}$ / cycle, TFSAM: -0.068 mAh g$^{-1}$ / cycle) it becomes clear that, despite the stabilization, TFSI cells still lose capacity at a faster rate than TFSAM. For a more detailed insight into the cycling performance, the voltage profiles are shown in Figure 7b. In the first cycles, the Ohmic drop (ΔV) is higher for TFSAM-based electrolytes (see SI, table S2 and S3 for exact values of ΔV and equivalent series resistance (ESR)), which is coherent with the higher initial internal resistance of the electrolyte. However, it can be seen that, due to the faster Li$^+$ transport, the TFSAM cells reach the LFP plateau faster with less sloped profiles and smaller hysteresis between charge and discharge (*i.e.* higher energy efficiencies). Over the course of cycling, ΔV and ESR increase for both electrolyte systems. Interestingly, this increase is faster for TFSI, leading to higher resistance values for TFSI in the long run (*e.g.*, ESR at the 300$^{th}$ cycle: 808 mΩ cm$^2$ for TFSI, and 707 mΩ cm$^2$ for TFSAM respectively).

To highlight the effect of the faster Li$^+$ transport, the discharge rate was increased to outline the influence of this increased Li$^+$ transport at higher discharge currents. The evolution of capacity (Figure 7c) is, for the lower C-rates, rather similar for both systems. However, at the highest rate (2C) the advantage of TFSAM is quite clear as the TFSAM-based cells deliver considerably higher discharge capacities. A closer look to the voltage profiles (Figure 7d) reveals that the previously observed behavior becomes more obvious as the current increases: In spite of a slightly higher ohmic drop and ESR (see SI), TFSAM reaches the LFP plateau faster with less sloped discharge curves and higher capacities at high rates.

## 4 Conclusion

We propose in this study the use of strongly Li$^+$-coordinating anions to accelerate the lithium ion transport in ternary PEO-based polymer electrolytes. The unilateral and strong Li–TFSAM interaction seen for the liquid binary IL-based electrolyte emerges as highly beneficial in combination with a PEO host matrix. Whereas the weakly coordinating TFSI might constitute the better anion choice in conventional liquid electrolyte formulations, it concedes the lithium ions to the rather immobile polymer, which entangles around the Li$^+$ resulting in a strong coupling of the lithium dynamics to its segmental mobility, which result in slow lithium transport. Through using a strongly lithium-coordinating anion like TFSAM, MD simulations revealed that it is not only possible to strip the lithium ions from the PEO structure, but also to re-couple their dynamics from the polymer matrix to the anion. The experimental results, in turn, indicate a substantial enhancement of the lithium-ion-carried conductivity due to this structural and dynamical shifting: With TFSAM (compared to TFSI), $t_{Li^+}$ was increased by the factor six which leads to a tripling of $\sigma_{Li^+}$ despite a lower overall conductivity.

Finally, we demonstrated that the principle of coordinating anions in TSPEs can be applied in LMPB cells. In LFP‖Li metal cells, TFSAM shows a clear beneficial influence on the capacity retention (after 300 cycles: TFSAM: 86%, TFSI: 82%). Furthermore, the accelerated Li$^+$ transport was visible in the flatter voltage profiles and higher capacities at high rates (2C).

These results recommend a rethinking of the role of coordinating anions in ternary polymer electrolytes and might lead to numerous future advances in this field of research, as many anions have been proposed over the years and did not make the cut for a competitive use as either liquid electrolyte (in organic solvent-based or IL-based electrolytes) nor in 'dry' SPEs due to insufficient salt dissociation.

## 5 Acknowledgement

The research leading to these results has been partially funded by the "GrEEn" project funded by the ministry of economy, innovation, digitalization, and energy of the state North Rhine-Westphalia, Germany (ETN project number: 313-W044A). Analysis and simulations have been performed on the computing cluster PALMA2 at the University of Münster.## 6 References

[1]  J.R. Nair, L. Imholt, G. Brunklaus, M. Winter, Lithium Metal Polymer Electrolyte Batteries: Opportunities and Challenges, Electrochem. Soc. Interface. 28 (2019) 55–61. https://doi.org/10.1149/2.F05192if.

[2]  C. Iojoiu, E. Paillard, Solid-State Batteries with Polymer Electrolytes, in: Encycl. Electrochem., 2020: pp. 1–49. https://doi.org/10.1002/9783527610426.bard110014.

[3]  I.E. Kelly, J.R. Owen, B.C.H. Steele, Poly(ethylene oxide) electrolytes for operation at near room temperature, J. Power Sources. 14 (1985) 13–21. https://doi.org/10.1016/0378-7753(85)88004-6.

# Are Weakly Coordinating Anions really the Holy Grail of Ternary Solid Polymer Plasticized by Ionic Liquids? Coordination Anions to the Rescue of the Lithium Ion Mobility

*Jan-Philipp Hoffknecht, Alina Wettstein, Jaschar Atik, Christian Krause, Johannes Thienenkamp, Gunther Brunklaus, Martin Winter, Diddo Diddens, Andreas Heuer, Elie Paillard*

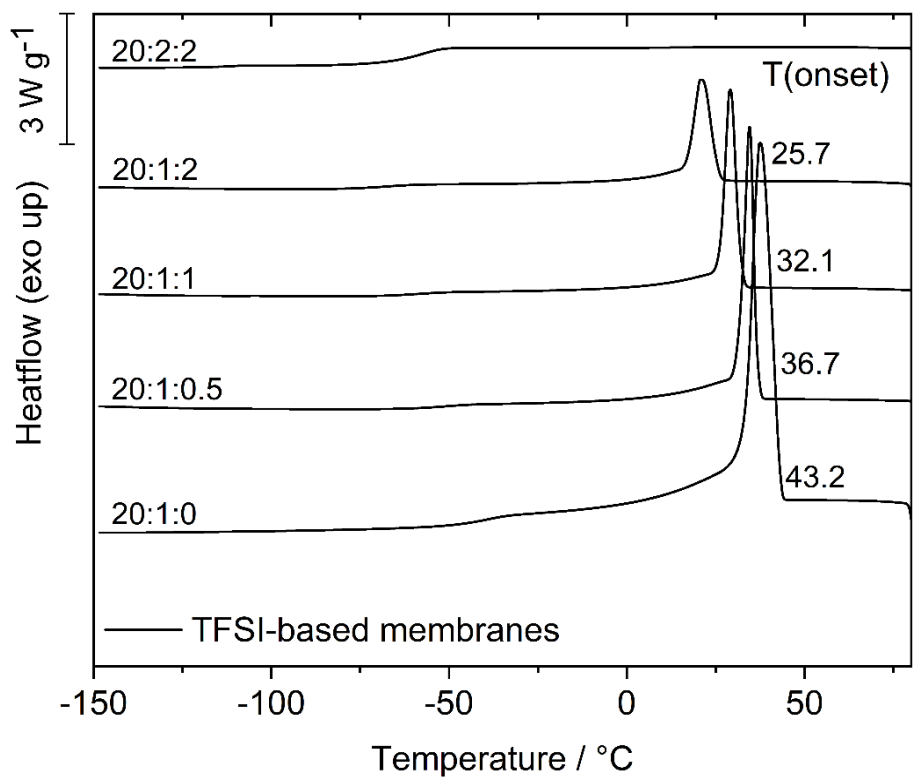

Figure S1: DSC cooling curves (after the shown heat ramp) of TFSI-based ternary electrolytes with different PEO:salt:IL ratios; onset point of the crystallization peaks are marked (analysed with Universal Analysis 2000 from TA Instruments).

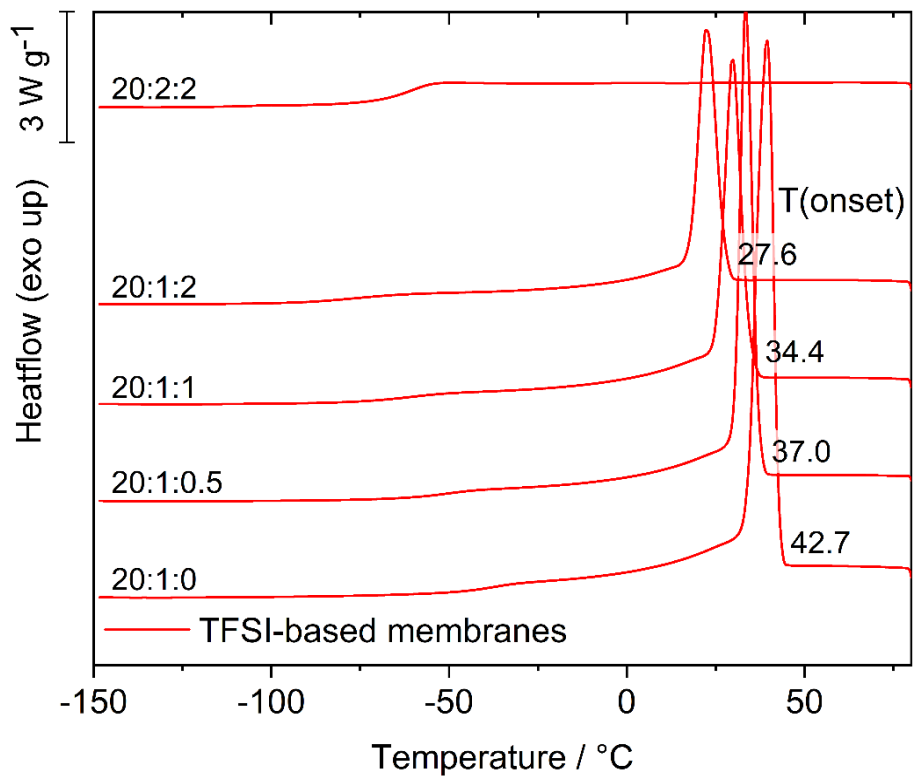

Figure S2: DSC cooling curves (after the shown heat ramp) of TFSAM-based ternary electrolytes with different PEO:salt:IL ratios; onset point of the crystallization peaks are marked (analysed with Universal Analysis 2000 from TA Instruments).

Table S1: Thermal behavior (glass transition, cold crystallization, melting) of TFSAM and TFSI based ternary electrolytes with different PEO:salt:IL ratios analysed from the heat ramps (figure S1 and S2).

| Membrane | Tg / °C | Cold Crystallization | | | Melting | | |
|---|---|---|---|---|---|---|---|
| | | T(onset) / °C | T(peak) / °C | Hc / J g$^{-1}$ | T(onset) / °C | T(peak) / °C | Hm / J g$^{-1}$ |
| TFSAM 20:2:2 | -54.1 | 1.6 | 20.8 | 5.5 | 36.4 | 43.3 | 4.6 |
| TFSAM 20:1:2 | -59.7 | -20.7 | -8.6 | 40.0 | 37.4 | 52.7 | 47.3 |
| TFSAM 20:1:1 | -53.7 | -22.1 | -14.4 | 7.7 | 40.4 | 54.0 | 60.4 |
| TFSAM 20:1:0.5 | -43.7 | -28.2 | -22.8 | 0.8 | 44.0 | 55.9 | 61.1 |
| TFSAM 20:1:0 | -30.3 | | no cold crystallization | | 51.7 | 60.5 | 89.1 |
| | | | | | | | |
| TFSI 20:2:2 | -51.7 | 0.5 | 20.8 | 4.4 | 37.3 | 43.4 | 37.3 |
| TFSI 20:1:2 | -56.2 | -17.1 | -7.8 | 17.1 | 35.7 | 49.4 | 25.3 |
| TFSI 20:1:1 | -50.3 | -25.0 | -12.0 | 1.3 | 39.4 | 51.8 | 42.3 |
| TFSI 20:1:0.5 | -43.9 | -22.3 | -2.3 | 2.0 | 42.5 | 54.2 | 49.4 |
| TFSI 20:1:0 | -31.8 | | no cold crystallization | | 50.4 | 59.3 | 69.6 |

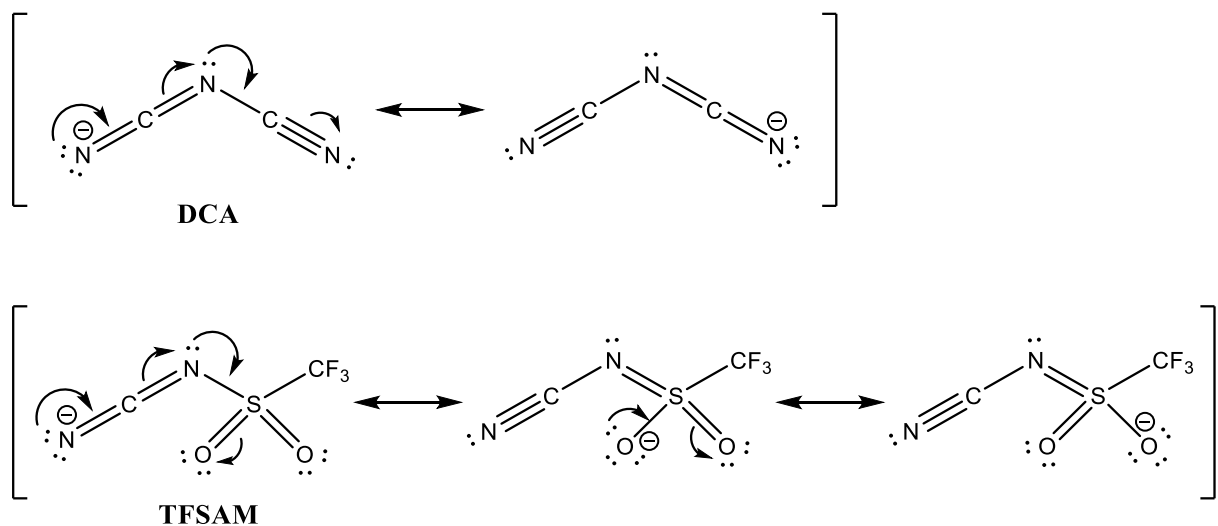

Figure S3: Mesomeric Structures of the DCA and the TFSAM anion.

Table S2: Ohmic drop and Equivalent Series Resistance of LFP‖Li metal cells with TFSI-based and TFSAM-based ternary solid polymer electrolytes; long-term cycling.

| Cycle Number | Charge / Discharge Rate | Ohmic Drop (ΔV) / V | | Equivalent Series Resistance / mΩ cm$^2$ | |
| --- | --- | --- | --- | --- | --- |
| | | TFSI | TFSAM | TFSI | TFSAM |
| 1 | 0.1 C / 0.1 C | 0,0163 | 0,0166 | 543 | 551 |
| 5 | 0.5 C / 0.5 C | 0,0823 | 0,0793 | 548 | 527 |
| 50 | 0.5 C / 0.5 C | 0,0925 | 0,0851 | 615 | 566 |
| 100 | 0.5 C / 0.5 C | 0,0992 | 0,0897 | 660 | 597 |
| 200 | 0.5 C / 0.5 C | 0,1112 | 0,0975 | 740 | 649 |
| 300 | 0.5 C / 0.5 C | 0,1215 | 0,1063 | 808 | 707 |

Table S3: Ohmic drop and Equivalent Series Resistance of LFP‖Li metal cells with TFSI-based and TFSAM-based ternary solid polymer electrolytes; rate performance test.

| Cycle Number | Charge / Discharge Rate | Ohmic Drop (ΔV) / V | | Equivalent Series Resistance / mΩ cm$^2$ | |
| --- | --- | --- | --- | --- | --- |
| | | TFSI | TFSAM | TFSI | TFSAM |
| 1 | 0.1 C / 0.05 C | 0,0123 | 0,0153 | 547 | 679 |
| 2 | 0.1 C / 0.05 C | 0,0120 | 0,0152 | 532 | 673 |
| 3 | 0.1 C / 0.05 C | 0,0125 | 0,0149 | 553 | 659 |
| 4 | 0.1 C / 0.1 C | 0,0147 | 0,0178 | 490 | 591 |
| 5 | 0.1 C / 0.1 C | 0,0153 | 0,0181 | 508 | 602 |
| 6 | 0.1 C / 0.1 C | 0,0152 | 0,0183 | 506 | 608 |
| 7 | 0.1 C / 0.2 C | 0,0199 | 0,0242 | 441 | 537 |
| 8 | 0.1 C / 0.2 C | 0,0201 | 0,0242 | 445 | 537 |
| 9 | 0.1 C / 0.2 C | 0,0201 | 0,0245 | 445 | 544 |
| 10 | 0.1 C / 0.5 C | 0,0349 | 0,0407 | 387 | 452 |
| 11 | 0.1 C / 0.5 C | 0,0351 | 0,0406 | 390 | 450 |
| 12 | 0.1 C / 0.5 C | 0,0351 | 0,0410 | 389 | 455 |
| 13 | 0.1 C / 1 C | 0,0609 | 0,0718 | 369 | 434 |
| 14 | 0.1 C / 1 C | 0,0619 | 0,0724 | 374 | 438 |
| 15 | 0.1 C / 1 C | 0,0611 | 0,0726 | 370 | 439 |
| 16 | 0.1 C / 2 C | 0,1098 | 0,2441 | 348 | 404 |
| 17 | 0.1 C / 2 C | 0,1105 | 0,1323 | 350 | 419 |
| 18 | 0.1 C / 2 C | 0,1110 | 0,1341 | 352 | 425 |